\documentclass[a4paper,12pt,twoside]{article}

\usepackage{t1enc}
\usepackage[utf8]{inputenc}

\usepackage{natbib}

\usepackage[bindingoffset=1cm,textheight=22cm,hdivide={2cm,*,2cm},vdivide={3.8cm,22cm,*}]{geometry}

\begin{document}

% \pagestyle{empty}                            % cimoldalhoz
% \frenchspacing
% \setstretch{1.3}                             % ezt kell kivenni takarékos nyomtatáshoz
\pagenumbering{arabic}

%%%%%%%%%%%%%%%%%%%%%%%%%%%%%%%%%%%%%%% CÍMLAP

{\centering
{\large
\ \\
Konkoly Observatory\\
Occasional Technical Notes\\
Number 15.
\vskip4mm
\textsc{\textbf{LCfit,}}
\vskip1mm
\textbf{a harmonic-function fitting program}
\vskip4mm
% \textsc{Sódor Ádám}
Ádám Sódor}
\vskip1mm
Konkoly Observatory\\
Research Centre for Astronomy and Earth Sciences,\\ Hungarian Academy of Sciences, Budapest, Hungary
\vskip2mm
29 May 2012

}

\section{introduction}

In this note I announce and introduce the program {\sc LCfit} developed for fitting harmonic functions to a data set, particularly to time-series data. {\sc LCfit} stands for Linear Combination fitting.

The program has linear and non-linear fitting modes and uses an implementation of the Levenberg--Marquardt (LM) least squares fitting algorithm written by \cite{lourakis} for optimizing the parameters. The user gives a data set, base frequencies and linear-combination terms as input. In linear mode, {\sc LCfit} fits the phases and amplitudes of the linear-combination terms to the data. In non-linear mode, the base frequencies are also fitted using the user input as initial values for these parameters. The result of the fit is given in different forms. The formal uncertainties of the fitted parameters, calculated from the covariance matrix of the results are also given in the output. The program is written in C++. The source code is freely distributed under the terms of the GNU General Public License\footnote{\tt http://www.gnu.org/copyleft/gpl.html} and is available from the author.\footnote{Web page: {\tt http://konkoly.hu/staff/sodor/lcfit.html}}$^,\, $\footnote{e-mail: {\tt sodor *at* konkoly *dot* hu}}

\section{usage}

Brief instructions on the usage is written on the standard output if the program is run without sufficient number of command-line arguments:

\newpage

{\scriptsize
\begin{verbatim}
    Usage: lcfit [options] data fitdef [pw_output]

      where: 
        options:
          -g        : gnuplot output
          -v        : verbose
          -c {file} : mufran .coe output to file {file} (if used, -c option must be the last one)
        data        : file containing the light curve (time series)
        fitdef      : file describing the frequencies and lincombs to fit.
        pw_output   : file to write prewhitened output. Will be overwritten!

\end{verbatim}
}

The data file is a simple list of whitespace separated \verb+time value+ pairs. Empty lines and lines beginning with '{\tt \#}' are ignored. The {\tt fitdef} file format is introduced in the next section. If the {\tt pw\_output} command-line argument is given, then a time series prewhitened with the fitted solution is written into that file.

\subsection{The {\tt fitdef} file}
\label{sect:fitdef}
The harmonic functions to fit, as well as some parameters of the fitting algorithm are given in the {\tt fitdef} input file, which has the following syntax:
\vskip3mm
{\scriptsize
\begin{verbatim}

    [fixfreq]

    sid = 1.0024

    [varfreq]                                                                                                                                                                                                                                                                         

    f1 = 2.1453
    f2 = 0.0149                                                                                                                                                                                                                                                                         

    [lincomb]                                                                                                                                                                                                                                                                         

    sid
    f2

    f1                                                                                                                                                                                                                                                                                 
    2*f1

    f1 + f2
    3*f1 - 2*f2

    [LMpar]                                                                                                                                                                                                                                                                           

    maxiter = 5000                                                                                                                                                                                                                                                                    
    #epsilon2 = 3e-7                                                                                                                                                                                                                                                                  
    #mu=1e-8                                                                                                                                                                                                                                                                          
    t0 = 54540.5
\end{verbatim}
}

\vskip3mm
The {\tt fitdef} file is divided into four sections by the headings: {\tt [fixfreq]}, {\tt [varfreq]}, {\tt [lincomb]}, and {\tt [LMpar]}. The section headings are case sensitive and their order is fixed. However, not each of them is mandatory. One of the frequency-definition sections ({\tt [fixfreq]} or {\tt [varfreq]}) and {\tt [LMpar]} might be omitted, but one of the frequency definition sections and {\tt [lincomb]} are mandatory. Empty lines and lines beginning with '{\tt \#}' mark are ignored, as well as whitespaces around the {\tt +}, {\tt -}, {\tt *} and {\tt =} marks and at the beginning and end of lines.

The base frequencies not to be fitted are given in the {\tt [fixfreq]} section in \verb+identifier = value+ format, one on each line. The frequency unit is the inverse of the unit of time used in the input {\tt data} time-series file. The frequencies given in the {\tt [varfreq]} section will be fitted by {\sc LCfit}. It is mandatory to give initial frequency values for the fitted frequencies. If at least one frequency is given in the {\tt [varfreq]} section, a non-linear fit will be calculated.

The actual Fourier terms to be fitted are given in the {\tt [lincomb]} section. Note that harmonic terms of the base frequencies will be fitted only when the given base frequency is listed here.

The last, optional section, {\tt [LMpar]} lists further parameters for the program: {\tt t0} is the epoch for which the output should correspond (its default value is the integer part of the time of the first data point of the input data file). The maximum number of iterations of the LM algorithm is given by {\tt maxiter} (its default value is 1500, which might be insufficient for a complex fit). The step-size for numeric difference calculations, which is set to its default by {\sc LCfit} according to the length of the data set in order to fit the frequencies appropriately, is given by {\tt delta}. The modification of this parameter by hand is usually not necessary. For completeness, we mention {\tt mu}, {\tt epsilon1}, {\tt epsilon2}, and {\tt epsilon3}, which are further LM parameters (for details we refer the reader to the documentation of the LM algorithm in \citealt{lourakis}). Their default values are almost always suitable, thus, it is seldom necessary to override them.

\subsection{Output}

The result of the fit is written to the standard output in the following format:

{\scriptsize
\begin{verbatim}
Final parameters:
-----------------

Results for harmonic curve fit:
LM stopped after 70 iterations and 95 function evaluations.                                                                                                                                                                                                                       
Exit reason: convergence (2 - stopped by small Dp)                                                                                                                                                                                                                                

Minimization info:                                                                                                                                                                                                                                                                
Final r.m.s. = 0.00984932   Reduced Final r.m.s. = 0.00986411

           sid =   1.0024000000; # (fix freq.)
            f1 =   2.1452983111; Df1             =   1.0000000000
            f2 =   0.0148972133; Df2             =   0.0000020287

  0.00011    0.00031    55490.00000 # A0 DA0 epoch

#N   A1        DA1       A2       DA2        A        DA        fi       Dfi       f            lincomb
001  0.001546  0.000002  1.001022 0.000442   1.001023 0.000441  1.569252 0.000885  1.0024000000 (+1*sid)
002 -0.003992  0.000394  0.000678 0.000040   0.004049 0.000376  2.973280 0.103134  0.0148972133 (+1*f2)
003 -0.473363  0.000497 -0.123707 0.000090   0.489261 0.000570  3.397212 0.001550  2.1452983111 (+1*f1)
004 -0.247322  0.000640  0.056327 0.000077   0.253655 0.000599  2.917664 0.003001  4.2905966221 (+2*f1)
005 -0.007902  0.913895 -0.003489 0.403028   0.008638 0.000546  3.557396 0.254601  2.1601955244 (+1*f1+1*f2)
006 -0.000214  0.540259 -0.000333 0.841428   0.000396 0.000654  4.141593 1.124066  6.4061005065 (+3*f1+-2*f2)
\end{verbatim}
}

In the first part, a summary can be read about the outcome of the fit.

The \verb+Minimization info+ section is self-explanatory.

The next section gives the results of the frequency fit. The example shows the output of a fit with both fixed ({\tt sid}) and fitted ({\tt f1}, {\tt f2}) freqencies, according to the example {\tt fitdef} file. The formal uncertainties are also given for the fitted frequencies.

After the frequency section follows a single line with the fitted zero-point of the data ({\tt A0}), its uncertainty ({\tt DA0}) and the {\tt epoch} or $t_0$ to which the Fourier parameters listed in the last section correspond. %The default epoch is the integer part of the time of the first data point of the input data file, which can be modified in the {\tt fitdef} file (see Sect.~\ref{sect:fitdef}).

The last section lists the Fourier parameters of the fitted linear-combination terms. The first column is a serial number of the term, followed by {\tt A1} and {\tt A2}, the amplitudes in cosine- and sine-term (or real and imaginary part) formalism, while {\tt A} and {\tt fi} are amplitudes and phases in the sine-amplitude and -phase formalism. The amplitudes are given in the same unit as used in the input file, while the phases are in radians. {\tt D$x$} denotes the uncertainty of $x$. The last two columns give the frequency of the term ({\tt f}) and the linear-combination expression ({\tt lincomb}).

In case the {\tt -g} ({\sc gnuplot} output) option is selected on the command line, then the results written to the standard output are formatted in a different way so that the output can directly be used as input for {\sc gnuplot} as in the following example:\\

{\footnotesize
\verb+$ lcfit -g lightcurve_file fitdef_file | gnuplot -persist+
\\
}
% (anything not meant for gnuplot are preceded by a comment mark (\#), the fitted function are given in gnuplot format, variables are set, a plot command is written, etc.)

As a result, gnuplot shows the data and the fitted curve for inspection.

An alternative output format can also be chosen with the \verb+-c {file}+ command line option. In this case, the results are not only printed on the standard output, but the Fourier parameters are also written into \verb+{file}+ in the Fourier coefficient file format of {\sc MuFrAn} \citep{mufran}.

In verbose mode, further information are printed during the run of {\sc LCfit} both on the standard error and on the standard output channels. If the {\tt pw\_output} command-line argument is given, then a time series prewhitened with the fitted solution is written into that file.

\end{document}